# AN IMPROVED SYSTEM FOR SENTENCE-LEVEL NOVELTY DETECTION IN TEXTUAL STREAMS

*Xinyu Fu[1], Eugene Ch'ng[1], Uwe Aickelin[2], Lanyun Zhang[1]*

[1]*International Doctoral Innovation Centre, The University of Nottingham, Ningbo, China*
[2]*School of Computer Science, The University of Nottingham, UK*
*Xinyu.FU@nottingham.edu.cn, Eugene Ch'ng@nottingham.edu.cn*
*Uwe.Aickelin@nottingham.ac.uk , Lanyun.ZHANG@nottingham.edu.cn*

## Abstract

Novelty detection in news events has long been a difficult problem. A number of models performed well on specific data streams but certain issues are far from being solved, particularly in large data streams from the WWW where unpredictability of new terms requires adaptation in the vector space model. We present a novel event detection system based on the Incremental Term Frequency-Inverse Document Frequency (TF-IDF) weighting incorporated with Locality Sensitive Hashing (LSH). Our system could efficiently and effectively adapt to the changes within the data streams of any new terms with continual updates to the vector space model. Regarding miss probability, our proposed novelty detection framework outperforms a recognised baseline system by approximately 16% when evaluating a benchmark dataset from Google News.

## 1 Introduction

In the recent decades, novelty detection or first story detection in social media analytics has been one of the trending research subjects due to the prevalence of large social networks such as Twitter and Sina Weibo. The literature has emphasised the importance of First Story Detection (FSD) as a variant of Topic Detection and Tracking (TDT). This is certainly true in the case of the recognised TDT-1 to TDT-5 competition and the TREC because a significant quantity of papers dealing with FSD have been published through the aforementioned two research institutions. FSD has emerged as a powerful technique for improving the accuracy and timing of news alert systems. In addition, FSD can play an important role in addressing the issue of Smart City monitoring, amongst which is the identification of the virality timing of an epidemic or hazards. FSD was first introduced with a pilot study for monitoring streaming news stories [1]. In this circumstance, the task was to assign each story with either novel or redundant.

The majority of past research in novelty detection has been focused on the time-complexity or the execution time of the system. This is exemplified in the work undertaken by [2]. According to [2], a constant time approach was taken to eliminate the time consuming operation caused by document to document comparisons. In traditional novelty detection frameworks, there has been a lack in the investigation of the changing dynamics of the streaming context. Hence the consequent novelty detection step often does not reflect the updated term weighting information. A possible consequence of using the traditional TF-IDF weighting is that incorrect labelling of novel or redundant terms on a document may occur. This is a major issue in this kind of Big Data research.

Here, we present a new FSD algorithm based on the TF-IDF scheme which takes newly introduced vocabularies into account. Based on such an idea, we project that the new algorithm will outperform the existing baseline novelty detection system. In other words, the incremental TF-IDF weighting approach yields a more accurate identification on the text streams in a novelty detection system when compared to the baseline. The evidence is shown by a standard evaluation applied on a benchmark dataset.

The structure of the article is as follows. Existing literature in relates to FSD is presented in section 2. In section 3, we demonstrate the methodology which includes an overview of the basic model, the current state-of-the-art FSD algorithm and our novel system based on the existing algorithm. Section 4 presents the results and section 5 concludes the research with future work.

## 2 Literature Review

In this subsection, we present the background of various techniques and approaches encompassing the field of novelty detection.

Novelty detection was firstly introduced in signal processing research which sought the identification of novel signals for processing. A report on methods for novelty detection can be acquired from the Signal Processing Journal by [3]. The report has two separate sections: statistical methods and neural networks approaches. Here, we present a summary of methods related to our work.

It was argued by Allan et al. [4] that there were different similarity measures, ranging from cosine distance measure to distributional language models. An example of similarity measure was given in which agglomerative clustering methods were extensively explored to determine if a news story is describing a new event or a redundant one as compared to the previously identified documents. Another example of this is the k-nearest neighbour approach, which calculates the distance of the query document to the *k* nearest neighbouring documents respectively and assigns the query document to the group label which has most neighbours out of *k*.

There has been a sheer amount of literature indicating that cosine similarity can outperform other language models and the KL divergence technique. For cosine similarity, previous studies have reported that two major variations are worth considering: max cosine similarity or 1-NN approach and mean cosine similarity. For max cosine similarity measure, the maximum value is selected from a group of document to document comparisons. However, for mean cosine similarity measurement, the mean value from each document to document comparison is output.

The methodologies mentioned in the literatures are computationally expensive due to the brute force document to document comparison. Petrovic et al. [2] approximate 1-NN with LSH. Some work has been carried out by Song et al. [5] to improve the efficiency of novelty detection systems by introducing a news indexing-tree. [6] presents a framework for online new event detection used in a real application. The approach used is called 1-NN approach and the framework focuses on improving system efficiency by reducing the number of saved documents using indices, parallel processing, and etc. As compared to prior research, our method increases the efficiency of novelty detection by avoiding the exhaustive comparisons present in 1-NN approach, this is described in the methodology section.

Previous studies have been primarily concerned with how a vocabulary can be individually weighted, and hence FSD systems generally implement static term weighting scheme. This can be seen in the case of the two well-known TDT systems, namely the UMass and the CMU system [2]. To the best of our knowledge, only few investigations have been done on appropriate yet efficient weighting of a token. One of these is the work by Brants et al. [7], the FSD system is a typical instance of incremental TF-IDF weighting framework, the work reported an incremental performance on the standard TDT3 and TDT4 datasets.

# 3 Methodology

In this section, the essential elements for building up a traditional FSD system are presented. Until recently, TF-IDF is the most popular technique for document representation and the core for term weighting for the New Event Detection task. Four participating systems evaluated in TDT-2002 uses it, in line with [7].

## 3.1 Vector Space Model

The traditional approach to new event detection or FSD is to represent documents as vectors in term space, where each dimension corresponds to a distinct token or term. If a term appears in the document, then its value in the vector is non-zero. Note queries are denoted as vectors as well as the documents. Each new document is then compared to the previous ones, and if it has similarity to the closest document (or centroid) below a certain threshold, the new document is declared as a First Story or new event. This approach is used in the UMass and the CMU system [4].

Several different rules for token weighting have been developed so far. The one we utilise for new event detection task is a popular and recognized term weighting scheme abbreviated as TF-IDF. In accordance with [7], all four participants in the TDT-2002 contest use TF-IDF for robust and efficient new event detection.

## 3.2 Pre-Processing

For pre-processing, we tokenise the data, remove stopwords, replace tokens by their stems [8], and generate inverse document frequencies vectors on a dynamic vector space model.

## 3.3 Term Frequency

Term frequency is defined as a value of the frequency of term $t_j$ in the document $D_i$.

## 3.4 Document Frequency

The document frequency is used to count the number of document contains term $t_j$.

## 3.5 Inverse Document Frequency

Inverse document frequency is expressed as *lg (d / df_j)* where *d* is the total number of documents in a collection or corpus and $df_j$ indicates the number of document contains term $t_j$ or the document frequency of term $t_j$.

## 3.6 Term Weighting

$$W_{i,j} = \frac{(\lg tf_{i,j} + 1.0) \times idf_j}{\sqrt{\sum_{j=1}^{t}[(\lg tf_{i,j} + 1.0) \times idf_j]^2}} \quad (1)$$

Term weighting is summarised in Equation 1. $W_{ij}$ denotes the weight of term *j* in the document numbered *i*. $tf_{ij}$ is defined as the term frequency of term *j* in the document *i*. Similarly, for $idf_j$, the inverse document frequency of term *j* is defined. Apparently an individual term is weighted over a map of all available terms in the model. That is why we have the expression in the denominator of Equation 1.

## 3.7 Similarity Calculation

Due to the fact that every document may have different mapping of terms and similarity calculation should eliminate the effect of document length, hence we applied a document length normalisation technique. For the document length normalisation, we use the cosine similarity measure as it has proven to be a standard way of measuring novelty score in novelty detection related implementations.

$$SC(Q, D_i) = \frac{\sum_{j=1}^{t} w_{q,j} d_{i,j}}{\sqrt{\sum_{j=1}^{t}(d_{i,j})^2 \sum_{j=1}^{t}(w_{q,j})^2}} \quad (2)$$

In Equation 2, the cosine similarity value of document $D_i$ and query Q is defined as the summation of the individual term weight $w_{qj}$ in the document $Q$ multiplying by $d_{ij}$ in the document $D_i$. The intermediate summed value is subsequently divided over the normalised document length of Q multiplying by the normalised length of $D_i$.

### 3.8 Locality Sensitive Hashing

It has been suggested that LSH is capable of improving computational time complexity on high dimensional data driven [2] nearest neighbor searching. Unlike the traditional 1-NN based framework, LSH does not require exhaustive computation of the distance of each pair of documents. LSH needs to only work out the cosine similarity values upon the trimmed hash buckets. The essence of LSH is that if two documents are far from each other in the original high-dimensional space, then the probability of them hashing into the same hash bucket is low. In other words, the possibility of the two given documents hashing into the same bucket is proportional to their distance in the original high-dimensional space.

Signature bits are utilised to calculate the Hamming distance between the two hashed documents. Random projection produces signature bits by separating hyper planes. More specifically, in a vector space, if the value of a target term weight is underneath the corresponding hyper plane, then we assign signature bit 0 to the corresponding cell, otherwise signature bit 1 is assigned.

We notice that by directly applying the traditional LSH algorithm on our FSD system, the result is a little bit worse than the expectation. We did some experiments and observed that one single hash bucket cannot effectively reflect the specificity of the training dataset. We need to introduce a new control variable $L$ for multiple hash tables' construction.

$$L = \log(\varphi - p^k) \quad (3)$$

where $p = \frac{\theta(x, y)}{\pi}$ and $\varphi$ is a preconfigured value of the probability of missing a nearest neighbor.

Each of the $L$ hash tables applies the same procedure as aforementioned and a union operator is deployed subsequently to get an overall average. By introducing the variable $L$, the cosine similarity measurement generates much more coherent result as compared with the singleton (a single hash table). Note that human annotators are used to manually classify each testing document and a ground truth collection is therefore available.

We present the pseudo code of the traditional algorithm which summarises our novelty detection algorithm (Table 1).

```
1.  Input threshold t
2.  for each document d in the collection
3.      do vector space model updating
4.      end do
5.  for each element in vector space model
6.      do LSH and put elements with identical hash values into buckets
7.      end do
8.  for each document s with same hash value of d'
        <- compare(testing document d', LSH)
9.      do min (distance (s, d'))
10.         if min distance < t
11.             assign d' as describing similar event with s
12.         else
13.             d' depicts a novel story
14.         endif
15.     end do
```

**Table 1. The traditional algorithm summarised in pseudo code**

### 3.9 Incremental TF-IDF Weighting Scheme

As the new testing documents come into the system with respect to time $t$, therefore it is worth considering the dynamic term weighting according to a time span. Note that the traditional static TF-IDF weighting approach does not take the new vocabularies from possible incoming data streams into account, which is very likely to cause inaccurate predictions. The real time streaming system usually has a significant document arriving rate, therefore our incremental TF-IDF weighting scheme would have a better representation of token weight.

Assume at time $t$-1, the term frequency is $tf_{(t-1)}$, and at time $t$, a bunch of new documents C arrived at the system with the term frequency $tf_{(C)}$; hence $tf_{(t)} = tf_{(t-1)} + tf_{(C)}$. Similarly, for inverse document frequency representation, the same philosophy applies, and we have the resembling expression $idf_{(t)} = idf_{(t-1)} + idf_{(C)}$ where $idf_{(t-1)}$ and $idf_{(t)}$ denote the inverse document frequency at time point $t$-1 and $t$ respectively, whereas $idf_{(C)}$ indicates inverse document frequency for the collection C.

Here, we illustrate our new algorithm presented using the pseudo code in Table 2. Note that the lines 4 to 9 is the successful implementation of the incremental TF-IDF weighting scheme which eliminates the problem of inaccurate term weighting that a static TF-IDF may have.

```
1.  For each document d in the collection
2.      do vector space model updating
3.      end do
4.  incremental tf-idf weighting
5.  for each new document in C
6.      do tf (t) = tf(t-1) + tf(C)
7.          idf (t) = idf(t-1) + idf(C)
8.          wij(t) = tf(t) * idf(t)
9.      end do
10. for each element in the updated vector space model
11.     do LSH and put elements with identical hash values into buckets
12.     end do
13. for each document s with same hash value of d'
        <- compare(testing document d', LSH)
14.     do min (distance (s, d'))
15.         if min distance < t
16.             assign d' as describing similar event with s
17.         else
18.             d' depicts a novel story
19.         endif
20.     end do
```

**Table 2. Pseudo code for our new algorithm**

### 3.10 Making a Decision

In order to decide whether a testing document $q$ is novel or redundant, it is iteratively compared to the $L$ hash tables of training documents. Within each iteration, a subset of documents $ds$ might be selected as describing similar event with document $q$, and we then union all $L$ iterations' results to achieve the overall result.

## 4 Experiments and Results

In this section we compare our new algorithm with the traditional algorithm, and present the benchmark dataset used for the experiments. The experiments were designed to quantify the improvement in error rates that can be achieved when term weighting is redefined as continuously adapting over time. Next, we present a discussion subsection which formulates a comparison between the previously mentioned result and the existing literature.

### 4.1 Presentation of Results

FSD systems are usually evaluated on a detection error trade-off (DET) curve [9]. We demonstrate a comparative evaluation of our promising algorithm with the baseline method based on DET curves.

### 4.2 Experimental Setup

Notice that a prerequisite for 5 fold or 10 fold cross validation is that the training document snippets have to be identified with either novel for describing a new event or redundant for depicting a previously seen event. However, the human annotation step may introduce another level of noise, therefore we reach a decision that no cross validation is performed in this particular experimental setting.

For the training documents, 500 snippets are chosen from the Google News dataset. Each of them is annotated and describes a new event different from each other.

We configure one group of testing Google News snippets with 1000 pieces of news within the group. We have 1000 predicted values which state either 0 or a document ID for each group of testing streams. Note that 0 indicates that this snippet describes a new event, otherwise a document ID of the previously seen event is assigned for this snippet.

We use Eclipse platform to perform the prediction step. For evaluation metrics, MATLAB is adopted. The experiments are run on a dual-core, 4GB RAM laptop computer.

### 4.3 Benchmark Dataset

News applications is generally used to evaluate FSD in textual contexts because this is the most popular and common type of text streams. There is a large volume of published studies utilising TDT test data for evaluation, in accordance with [4]. The TDT benchmark collection is poorly annotated: TDT5 covers 278,109 English news events but only 100 themes and approximately 4,500 labelled documents. Note TDT dataset is encoded in document-level. Our framework is instead designed for sentence-level novelty detection.

Another benchmark collection is the TREC dataset. This dataset is encoded in sentence-level. Both TDT and TREC dataset have the problem of manual annotation, which may have the issue of subjectivity.

However, in the case of Google News dataset, a reliable ground truth collection is pre-installed which eliminates the problem of human annotator as the leading decision maker. According to [3], Google News dataset is almost a machine annotated collection of news events which can be treated as a good benchmark for FSD performance metrics. The dataset covers news articles from the category "Technology" published within the time frame of July 12 to August 12, 2012. Over 60% of news data in the dataset is annotated, which introduces no bias compared to the full stream, regardless of the cluster size or cluster overlap.

### 4.4 Evaluation Metrics

In the literatures, the performance of a FSD algorithm is defined in accordance with miss rate and false alarm error rate. Allan et al. [10] develop a framework for the evaluation of TDT tasks where missed detection rate is the percentage of documents which should have been categorised as novel (but were not) to the total amount of documents that indicate as new and false alarm rate is the ratio of documents that mistakenly identified as novel to the total number of documents categorized as new. A variation of ROC curves, detection error trade-off (DET) can be used to demonstrate the trade-off between miss probability and false alarms. On the x-axis is the miss rate and on the y-axis is the false alarm rate. A system is considered to perform best when it has its curve towards the lower-left of the graph. The axes of the DET curve are on a Gaussian scale [11].

For DET curve plotting, first of all we need to introduce a normalized detection cost function

$$(C_{Det})_{Norm} = \frac{C_{Miss} \cdot P_{Miss} \cdot P_{t\arg et} + C_{FA} \cdot P_{FA} \cdot P_{nontarget}}{\min(C_{Miss} \cdot P_{t\arg et}, C_{FA} \cdot P_{nontarget})} \quad (4)$$

where $C_{Miss}$ is the cost of missing a new event, $P_{Miss}$ is the probability of missing a new event, $P_{t\arg et}$ is a priori probability of missing a new story, $C_{FA}$ is the cost of a false alarm, $P_{FA}$ is the probability of a false alarm, and $P_{nontarget}$ is the probability of seeing an old event [7]. A perfect system would score 0 in the normalized detection cost function. A naïve system which is always yielding yes or no scores 1.

### 4.5 Results

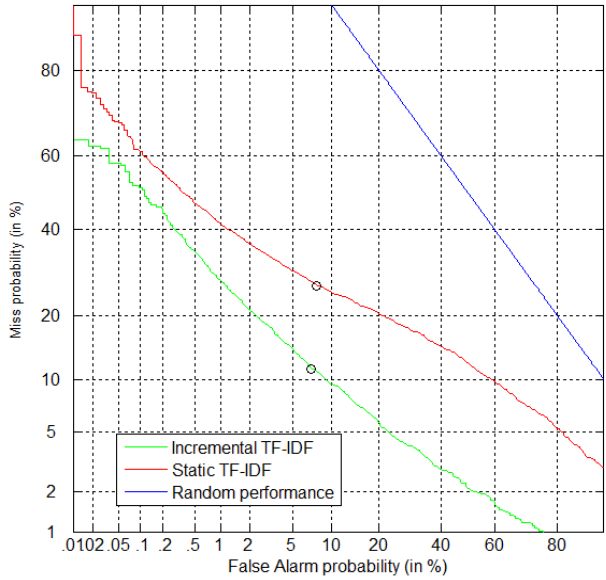

**Figure 3. DET curves for incremental TF-IDF weighting scheme, static TF-IDF respectively**

As shown by Figure 3, the DET curve for the incremental TF-IDF weighting scheme is below the DET curve for the traditional static TF-IDF weighting approach, which means the incremental TF-IDF approach performs better than the static TF-IDF approach. Obviously, for a fully random classification, the DET curve is a straight line connecting (0, 1) to (1, 0).

Notice the two small circles on the DET curves for the incremental TF-IDF and static TF-IDF are the lowest cost points for each scheme respectively. The best threshold for cost is directly related to miss probability and false alarm rate of a novelty detection system. Table 3 and 4 below describe the correspondent false alarm rate and missed detection probability for the static TF-IDF weighting scheme and incremental TF-IDF weighting approach respectively. It is evident from the following tables that the system applied with the incremental TF-IDF weighting scheme better classifies testing streams.

| Group | Static TF-IDF | Incremental TF-IDF |
|---|---|---|
| A | 8.01% | 7.20% |

**Table 3. False alarm probability for the testing set**

| Group | Static TF-IDF | Incremental TF-IDF |
|---|---|---|
| A | 27.1% | 11.6% |

**Table 4. Miss probability for the testing set**

Although the execution time of the incremental TF-IDF based new event detection system is a bit worse than the static one, the improvement of the performance outweighs the loss of the system operational time.

### 4.6 Discussion

An observation of the error rates extracted from Table 3 and Table 4 reveals that the incremental TF-IDF weighting scheme may exert a positive influence on reducing the error rate, including both the false alarm rate and missed detection possibility. There was a striking difference between the two weighting schemes as the incremental TF-IDF weighting scheme makes use of dynamic vector space model with respect to time whereas the static one only constructs a static term weighting structure for only the training streams.

In reference to the previous relevant literature, the state of the art novelty detection system could perform relatively well in some specific contextual streams, for example the TDT context. However, the general system performance has not been systematically reviewed yet. Hence, a lot of first story detection system is still context specific, that is why we are not capable of presenting a wealth of benchmark datasets for a complete system performance measurement. Note also our system focuses on the sentence-level first story detection as for the document-level new event detection, often the way of identifying new stories is significantly disparate.

In general, therefore, it seems that the source of evaluation datasets is still a limiting factor for improving the accuracy of the novelty detection systems. Being limited to the size of the evaluation dataset, this study lacks the experience of very large scale and real time streaming texts. As now the research community has entered the era of Big Data, this problem might be eased by the future investigations on collecting useful textual data.

## 5 Conclusion

In this paper, we provided the argument that the incremental TF-IDF weighting scheme is very likely to improve the accuracy of novelty detection system. Our experiment on the Google News dataset and the evaluation of the results confirmed our argument. We have noted from both the aforementioned results that to some extent, the training data may affect the predicted results. Additionally, the large and growing body of literature insists that novelty detection is not independent from the data streams available. Further work is required to establish the viability of our new algorithm. As it stands, there is room for further progress in improving the general novelty detection algorithm performance over various datasets, and, at the time of authorship, evaluations are being conducted using those datasets.

# 6 Acknowledgement

The author acknowledges the financial support from the International Doctoral Innovation Centre, Ningbo Education Bureau, Ningbo Science and Technology Bureau, China's MoST and The University of Nottingham. The project is partially supported by NBSTB Project 2012B10055.